\documentclass[reprint,twocolumn,superscriptaddress,showpacs,nofootinbib,notitlepage]{revtex4-1}

\usepackage{graphicx}
\usepackage{latexsym,amsmath,amssymb,lmodern,float,url}
\usepackage{natbib}
\usepackage{color}
\usepackage{microtype}
\usepackage{slashed}
\usepackage{multirow}
\usepackage{comment}
\usepackage{ulem}
\usepackage[colorlinks=true,backref=false, linktocpage=true,
citecolor=blue,urlcolor=blue,linkcolor=blue,pdfpagemode=UseOutlines]{hyperref}

\hypersetup{%
  bookmarksnumbered=true,
  pdftitle = {},
  pdfsubject = {},
  pdfauthor = {},
  pdfkeywords = {}
}

\newcommand{\mmax}{m_{\rm max}}

\renewcommand{\d}{\mathrm d}

\begin{document}
\title{The thermodynamics of large-$N$ QCD and the nature of metastable phases}
\author{Thomas D. Cohen}
\email{cohen@umd.edu}
\affiliation{Department of Physics, University of Maryland, College Park, Maryland 20742, USA}
\author{Scott Lawrence}
\email{srl@umd.edu}
\affiliation{Department of Physics, University of Maryland, College Park, Maryland 20742, USA}
\author{Yukari Yamauchi}
\email{yyukari@umd.edu}
\affiliation{Department of Physics, University of Maryland, College Park, Maryland 20742, USA}
\date{\today}

\begin{abstract}
In the limit of a large number of colors ($N$), both Yang-Mills and quantum chromodynamics are expected to have a first-order phase transition separating a confined hadronic phase and a deconfined plasma phase.  One aspect of this separation is that at large $N$, one can unambiguously identify a plasma regime that is strongly coupled. The existence of a first-order transition suggests that the hadronic phase can be superheated and the plasma phase supercooled. The supercooled deconfined plasma present at large $N$, if it exists, has the remarkable property that it has negative absolute pressure -- {\it i.e.}\ a pressure below that of the vacuum.  For energy densities of order unity in a $1/N$ expansion but beyond the endpoint of the hadronic superheated phase, a description of  homogeneous matter composed of ordinary hadrons with masses of order unity in a $1/N$ expansion can exist, and acts as though it has a temperature of $T_H$ in order unity.  However, the connection between the canonical and microcanonical descriptions breaks down and the system cannot fully  equilibrate as $N \rightarrow \infty$. Rather, in a hadronic description, energy is pushed to hadrons with masses that are arbitrarily large. The thermodynamic limit of large volumes becomes subtle for such systems: the energy density is no longer intensive. These conclusions follow provided that standard large-$N$ scaling rules hold, the system at large $N$ undergoes a generic first-order phase transition between the hadronic and plasma phases and that the mesons and glueballs follow a Hagedorn-type spectrum.
\end{abstract}

\maketitle

\section{Introduction}
Quantum chromodynamics (QCD) at finite temperature and zero chemical potential consists of a confined hadronic regime and a deconfined plasma regime, which are connected by a crossover~\cite{Aoki:2006we}. In this crossover regime, the medium is neither unambiguously hadronic nor unambiguously a plasma, and the physical description is not straightforward. The nature of matter near the crossover temperature can be probed with heavy-ion experiments. Phenomenological models of heavy-ion collisions have been studied to describe the collisions and have successfully explained many aspects of the experiments~\cite{Andronic:2017pug,Noronha-Hostler:2019nve,Schenke:2010nt,Gale:2013da}; however, the assumptions underlying these models, particularly in the crossover regime, are not always consistent with experiment~\cite{Cai:2019jtk}. 

Many of the awkward ambiguities of the crossover regime vanish in the large-$N$ limit of QCD~\cite{tHooft:1973alw,tHooft:1974pnl,Witten:1979kh}.  At large $N$ the hadronic and plasma regimes are expected to become unambiguously distinct phases separated by a well-defined phase transition. The clean separation of two phases arises due to different characteristic scalings with $N$: the energy density of the hadronic phases scales as $N^0$, whereas that of the plasma phases scales as $N^2$~\cite{tHooft:1973alw,tHooft:1974pnl,Witten:1979kh}. Thus, they cannot be smoothly connected --- at least at infinite $N$.  Moreover, there is strong reason to believe that the transition between the phases is first-order at sufficiently large $N$. We expect that the thermodynamics of large-$N$ QCD becomes equivalent to that of $SU(N)$ Yang-Mills theory due to the suppression of quark loops. Yang-Mills with $N=3$ is known to have a first-order transition~\cite{Yaffe:1982qf}, and lattice simulations suggest that the first-order transition persists at larger $N$~\cite{Lucini:2003zr,Lucini:2002ku,Lucini:2012gg,Panero:2009tv}, with a latent heat that appears to grow with $N^2$.  This behavior is precisely what one would expect if the first order transition persists up to infinite $N$. In the rest of the discussion in this paper, we will make the standard assumption that a first-order phase transition exists between the hadronic phase and the plasma phase in large $N$ QCD.  

With a first-order transition, the structure of the phase diagram becomes cleaner. The temperature of a homogeneous medium naively determines the phase of that medium --- if the medium's temperature is lower than the confinement temperature, the medium is in the hadronic phase and above it is in the plasma phase.  However, as a feature of the first-order transition, a hadronic medium can be superheated and a plasma supercooled, up to endpoints of those metastable regimes.   Apart from the assumption that a first-order transition persists in the $N \rightarrow \infty $ limit, the analysis in this paper will also assume it is of the generic type, with exactly two stable phases, and exactly two metastable phases. 

The metastable regimes are globally unstable homogeneous phases.  While the free energy would be reduced if the system went into an inhomogeneous mixed phase at the transition temperature, the system is locally stable against small amplitude fluctuations in energy density over large volumes.  Since large amplitude  fluctuations over large regions are exponentially unlikely, such metastable regimes can be very long-lived --- at least in the absence of external perturbations introducing spatially large fluctuations that could induce transitions to an inhomogeneous regime.

The metastable superheated (supercooled) phase does not extend to arbitrarily high (low) temperatures. Beyond  endpoints of the metastable regimes, the homogeneous medium becomes locally unstable --- any spatial fluctuation over a large region, no matter how small the amplitude, will grow exponentially.  Since the existence of such random fluctuations is a necessary feature of thermal systems, the homogeneous system will decay towards an inhomogeneous one, eventually yielding a mixed phase at the transition temperature.

This paper focuses on the thermodynamic features of these metastable and unstable regimes in the large-$N$ limit.  Much of the analysis will be from the perspective of the microcanonical ensemble in which the key functional relation is the entropy as a function of energy. However for certain parts of the analysis it is sensible to use a canonical description in which the free energy as a function of temperature is the key relation.  Note there is no distinction between the canonical and grand canonical ensembles for this system as we are taking all chemical potentials to be zero.     
 
Most of the analysis in this paper is done for large-$N$ QCD in the absence of any chemical potentials and with the standard version of the large-$N$ limit with quarks assumed to be in the fundamental representation of color.  The qualitative conclusions of this analysis hold equally well for a variety of large-$N$ gauge theories, including pure gauge (Yang-Mills). Numerical studies of such theories on a lattice may demonstrate these thermodynamic features of metastable and unstable regimes.
 
Although large-$N$ QCD is generally expected to behave differently from nature with $N=3$, the thermodynamic aspects of such metastable and unstable phases ought to be of interest for a better understanding of thermodynamics of gauge theories. Moreover, it is worth noting that for some observables the large-$N$ world is a recognizable caricature of the physical world with $N=3$ and can be used to make qualitative~\cite{Manohar:1998xv,Witten:1979kh} and sometimes semiquantitative predictions (for example baryon axial coupling constant ratios ~\cite{Dashen:1993jt}), of direct relevance to the physical world.  The thermodynamic issues at the heart of this paper are not of this type. Indeed, the behavior of QCD  at large $N$ with a first-order transition is qualitatively quite different from the $N=3$ world with a smooth crossover; the metastable phases on which this paper principally focuses do not even exist in the $N=3$ world.  Nevertheless, it is worth trying to understanding QCD thermodynamics at large $N$ as may give significant insight into QCD more generally. For example, as will be discussed in Sec.~\ref{gluon}, in the large-$N$ limit one can explicitly demonstrate that a strongly interacting plasma must exist --- this is in accord with the phenomenological understanding from the analysis of heavy ion collisions that a regime of strongly interacting plasma is formed~\cite{PhysRevLett.106.192301}.
 
The analysis in this paper depends on three assumptions:
 \begin{enumerate}
 \item Standard large-$N$ scaling rules hold for both properties of hadrons and for properties of a quark gluon plasma~\cite{tHooft:1973alw,tHooft:1974pnl,Witten:1979kh}. 
 \item As the large-$N$  and thermodynamic limits are approached, the  system, when fully equilibrated in a stable phase, has a single phase transition between a hadronic and plasma phases; the transition is generic first order and thus allows the system to superheat into a metastable hadronic regime and supercool into a metastable plasma phase (over a nonzero range of temperatures).  There are no other (meta)stable phases. \label{nonan}
 \item The mesons and glueballs have a Hagedorn spectrum~\cite{Hagedorn:1965st}.   This is a spectrum for which  $N^{\rm had}(m)$, the number of hadrons with mass less than $m$, at asymptotically high mass behaves as  $N^{\rm had}(m) \propto (m / T_H)^{-d} \exp(m/T_H)$;   $T_H$ is a parameter with dimensions of mass.  Moreover, $d$, the power in the subexponential prefactor is  assumed to greater than 7/2.  \label{Hagd}
 \end{enumerate}

Assumption \ref{nonan} is quite plausible in light of lattice studies of gauge theories at multiple values of $N$. The Hagedorn spectrum of Assumption~\ref{Hagd} is discussed in detail in Subsection~\ref{hagedorn}.
 
 
In Section~\ref{LargeN}, we review general features of first-order phase transitions from the perspective of the microcanonical ensemble; we also summarize the scaling of thermodynamic quantities in the two phases of large-$N$ QCD. Section~\ref{gluon} focuses on the plasma phase of large-$N$ QCD and establishes two facts: firstly, that the deconfined medium near the transition is a strongly coupled plasma, and secondly that if supercooled plasma exists over a finite temperature range in the large-$N$ limit, then it must achieve negative absolute pressure. It is not obvious how large $N$ needs to be to have a phase with negative pressure from our analysis. Numerical lattice simulations of Yang-Mills theories with large enough $N$ may be able to address this question. Section~\ref{hadron} focuses on the hadronic phase and the  regime after the endpoint of the metastable hadronic regime for the situation when $d>7/2$. We show that in the $N\rightarrow\infty$ limit, the energy density of homogeneous matter can be increased without limit while its temperature remains fixed to the Hagedorn temperature.  However, this requires the energy of the system to reside in hadronic states with arbitrarily high masses that may be beyond the regime of validity of standard large $N$ scaling rules for hadrons. We summarize these results and discuss open questions in section~\ref{Dis}.


\section{Microcanonical Ensemble and Large $N$}\label{LargeN}

\subsection{The microcanonical ensemble}
Since much of  the analysis in this paper is done using microscanonical reasoning, we begin by reviewing some elementary general features of first-order phase transitions as described via the  microcanonical ensemble~\cite{landau}.  

The key quantity in a microcanonical description of a finite volume system is  the entropy as a function of the energy $S(E)$, where the entropy of the system is the logarithm of the number of accessible states at that energy.  The temperature is defined as $1/T = S'(E)$.  We will focus on homogeneous  systems in the thermodynamic limit; therefore we will work with the intensive quantities such as the entropy and energy densities, $s=S/V$ and $\epsilon=E/V$.

\begin{figure}[t]
  \begin{center} 
    \includegraphics[width=9cm]{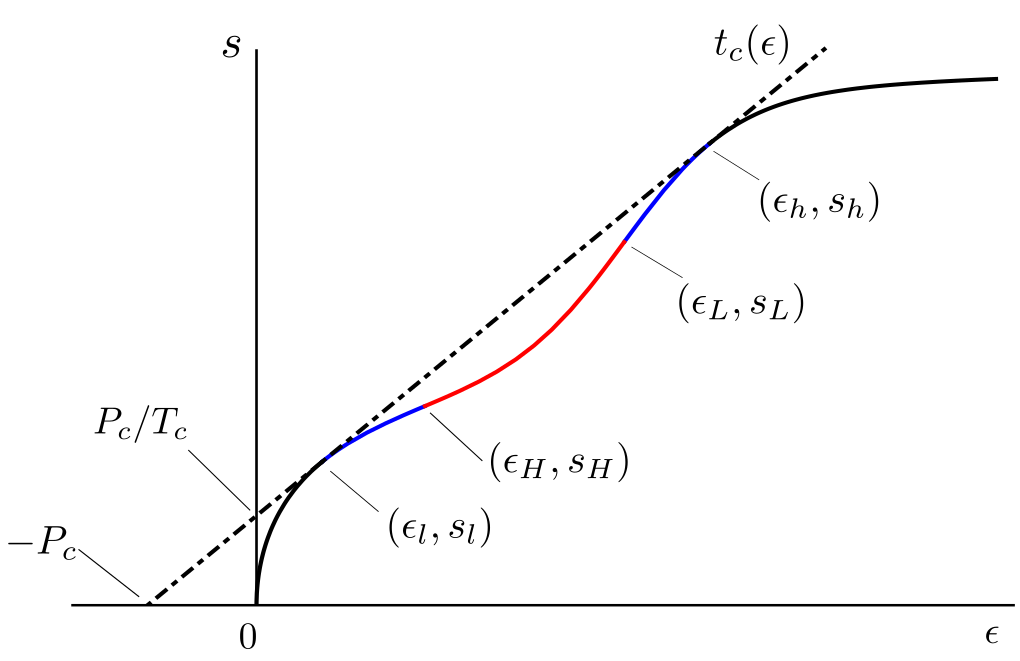}
    \caption{An entropy density-energy density curve for a system with a generic first-order phase transition.  As drawn, the curve is analytic everywhere. Despite this, the thermodynamics has nonanalaytic behavior. Note that, depending on details a system  with a generic first-order transition can have nonanalytic behavior of $s(\epsilon)$ at the points of inflection --- while the first derivative is continuous and the second derivative zero, higher derivatives can be discontinuous or divergent.
    \label{fig:mc-generic} } 
  \end{center}
\end{figure}

The pressure $P$ will play a central role in this work.  For a homogeneous system, the pressure (or equivalently, the negative of the free energy density) is
\begin{equation}
P(\epsilon)= -f = T s(\epsilon) -\epsilon\text.
\label{Pdef}
\end{equation}
Geometrically, both the temperature and pressure  for a system with energy density $\epsilon_0$ are given in terms of the line $t(\epsilon)$, which is defined as the  line in the $\epsilon$-$s$ plane  that is tangent to the curve $s(\epsilon)$ evaluated at the point $(\epsilon_0,s(\epsilon_0))$:
 \begin{equation}
t(\epsilon) = \frac{\epsilon + P(\epsilon_0)}{T(\epsilon_0)}
\text.
\end{equation}
The temperature is the multiplicative inverse of the slope of this line  while the pressure is given by the negative of the  $\epsilon$-intercept.  
Fig.~\ref{fig:mc-generic} shows the entropy density $s(\epsilon)$ as a function of energy density, for a homogeneous medium in a system which has a first-order phase transition at a critical temperature $T_c$. 

A key feature of a first-order transition is a region in which $s''(\epsilon)>0$ (see Fig.~\ref{fig:mc-generic}). This region corresponds to an absence of stable homogeneous configurations: at fixed energy, introducing an appropriate inhomogeneity will increase the entropy, and therefore inhomogeneous configurations are prefered. Critically, in the region where the entropy density is concave up, this is true even locally, so that even small amplitude fluctuations can `trigger' the instability (in contrast to metastable regimes).

The condition $s''(\epsilon)>0$ describes locally accessible instabilities, which can begin from arbitrarily small fluctuations. Homogeneous systems in the thermodynamic  limit can have regions which are globally unstable against the formation of inhomogeneities regardless of the sign of $s''(\epsilon)$. If two points $(\epsilon_1,s_1)$, $(\epsilon_2,s_2)$ lie on the homogeneous curve $s(\epsilon)$, and the line between them is above $s(\epsilon)$ everywhere in $\epsilon_1<\epsilon < \epsilon_2$, then a spatially separated system --- a mixed phase --- with part of the system with energy density $\epsilon_1$ and part with $\epsilon_2$ will have a higher entropy than the homogenous phase with the same average energy density.  Clearly, the optimal choice for a mixed phase is one which none of the system is concave upwards.  Thus, the equilibrium curve is constructed as the convex hull of all points on $s(\epsilon)$.  This is the famous Maxwell construction.

The illustrative curve shown in Fig.~\ref{fig:mc-generic} contains two phases --- as we expect happens for large $N$ QCD.  Note that the convex hull contains a line segment from $t_c(\epsilon)$ that is tangent to $s(\epsilon)$ at two points. These two points represent the properties of the two regions in the mixed phase.  Since the two points lie on the same line, the tangent lines share a common slope, and hence correspond to the same temperature, $T_c$, and  a common $\epsilon$-intercept and hence correspond to the same pressure.  For a medium at $T_c$, any energy density between $\epsilon_l$ and $\epsilon_h$ is achievable in equilibrium --- albeit in a mixed phase.  

In addition to the region $\epsilon_H < \epsilon < \epsilon_L$ where $s''(\epsilon) > 0$ and thus the system is locally unstable, there are two regions in Fig.~\ref{fig:mc-generic} where the medium is homogeneous and locally stable with $s''(\epsilon) < 0$, but nevertheless globally unstable. These metastable regions, defined by $\epsilon_l < \epsilon < \epsilon_H$ and $\epsilon_L < \epsilon < \epsilon_h$, and are respectively termed the superheated and supercooled phases. The two endpoints of metastable phases, $\epsilon_H$ and $\epsilon_L$, are the inflection points of $s(\epsilon)$ curve. Metastable phases do not decay until a sufficiently large thermal fluctuation appears, and their decay time is exponentially long in the size of the required fluctuation.

The superheated phase is smoothly connected to the phase defined by $T < T_c$, but nevertheless has a temperature above the critical temperature; correspondingly, the supercooled phase is smoothly connected to the hot phase with $T > T_c$, but nevertheless has a temperature below the critical temperature. 
To summerize, there are five regions on the homogeneous $s(\epsilon)$ curve:
\begin{enumerate}
    \item $ 0 < \epsilon < \epsilon_l$: cold, stable
    \item $\epsilon_l < \epsilon < \epsilon_H$: cold, metastable (superheated)
    \item $\epsilon_H < \epsilon < \epsilon_L$: locally unstable
    \item $\epsilon_L < \epsilon < \epsilon_h$: hot, metastable (supercooled)
    \item $\epsilon_h < \epsilon < \infty$: hot, stable
\end{enumerate}

\subsection{Large N Scaling}\label{largeN}

Let us now move to the specific case of large-$N$ QCD. The cold and hot phases correspond respectively to the hadronic phase and the quark-gluon plasma phase\footnote{In which gluons dominate, as there are $N^2-1$ gluon species.}. The transition between them is assumed, on the basis of strong lattice evidence, to be first-order~\cite{Lucini:2003zr,Lucini:2002ku,Lucini:2012gg,Panero:2009tv}. We denote the line of homogeneous medium for QCD with $N$ colors as $s_{N}(\epsilon)$, and will focus on properties that persist in the limit of large $N$.

The two phases possess different characteristic scalings of energy and entropy density with $N$. In the hadronic phase, both scale with $N^0$ (meaning that they have a finite limit as $N\rightarrow\infty$), while in the plasma phase, both scale with $N^2$, the number of species of gluon:
\begin{equation}
s_{N}(\epsilon) = \Lambda^3 \left \{ \begin{array}{l} N^0 \, \sigma\left(\frac{\epsilon}{N^0 \Lambda^4} \right) \; \;{\text{ for} \; } \epsilon \sim N^0\\    N^2 \, \Sigma\left(\frac{\epsilon}{N^2 \Lambda^4} \right) \; \; \;{\text{ for} \; } \epsilon \sim N^2 \end{array} \right. \text,
\label{sscale}\end{equation}
where $\Lambda$ is an arbitrary but fixed scale (chosen for convenience to be of order of the mass of a typical low-lying hadron), and $\sigma$  and $\Sigma$ are dimensionless functions characterizing the hadronic phase and  plasma phase respectively. When plotted on the $s$-$e$ plane, in the large-$N$ limit, the plasma phase scales out when focusing on the hadronic phase, and the hadronic phase shrinks into the origin when focusing on the plasma phase.

In the hadronic phase at low energy densities, the system becomes a noninteracting gas of  hadrons as the $N\rightarrow\infty$ limit is approached.   Since the baryon mass grows linearly with $N$~\cite{Witten:1979kh}, their contributions are exponentially suppressed at large $N$ for systems in thermal equilibrium, and can be ignored entirely. In contrast, the masses of mesons and glueballs are of order unity at large $N$~\cite{tHooft:1973alw,tHooft:1974pnl}.  The scattering amplitude for meson-meson  interactions is of order $O(N^{-1})$ while the  meson-glueball  and  glueball-glueball scattering amplitudes are of order $O(N^{-2})$~\cite{Witten:1979kh}.  Thus, at particle densties of order $unity$  (so as energy densities) the effects of interactions become negligible as $N \rightarrow \infty$.  Along with interaction strengths, hadronic widths also vanish:  $\Gamma_{\text{meson}} \sim O(N^{-1})$ and $\Gamma_{\text{glueball}}\sim O(N^{-2})$.   Finally, as shown by Witten~\cite{Witten:1979kh} the number of distinct glueballs and mesons becomes infinite in the large $N$ limit.  Thus, up to corrections of order $1/N_c$ the entropy density as a function of the  energy density when the energy density is of order $N^0$ is
\begin{equation}
s(\epsilon)= \sum_{k=1}^\infty (2 J_k+1) \, s^{\text{nbg}}(\epsilon;m_k).
\end{equation}
Here $s^{\text{nbg}}(\epsilon;m_k)$ is the entropy density of a noninteracting gas of  bosons of a single species with mass $m_k$ and $(2J_k+1)$ is the  spin degeneracy factor; effects of any flavor degeneracies  are included by treating these states as separate hadrons in the sum.

In the plasma phase at (asymptotically) high energy densities, the system is known to behave like a weakly interacting plasma of quarks and gluons, with gluons dominating the the large-$N$ limit.  The characteristic momentum scale for the gluons is $(\epsilon/N^2)^{1/4}$.  The $\beta$ function at large $N$ is for the `t Hooft coupling $\lambda=N g^2$, rather than $g^2$ itself~\cite{tHooft:1973alw};  the renormalization group evolution of $\lambda$  is independent of $N$ at large $N$.  Thus the characteristic momentum scale at which the theory becomes weakly coupled is independent of $N$, and therefore the corresponding characteristic energy density scales with $N^2$. Equivalently, the characteristic argument to $\Sigma$ is of order unity.

The previous argument implies that when the energy density is high enough for the perturbative expansion to be accurate --- which happens in the domain of $\epsilon \sim N^2$ --- $s_N$ scales as in Eq.~(\ref{sscale}). However, the plasma phase may extend to sufficiently low $\epsilon$ that the system is no longer perturbative.  Nevertheless, the scaling in Eq.~(\ref{sscale}) should hold; anything else would introduce an additional nonanalyticity in $s(\epsilon)$, in contradiction to the assumption of a signle generic first-order transition.

One important consequence of Eq.~(\ref{sscale}) is the behavior of the specific heat $c_V$:
\begin{equation}c_V = \frac{\partial \epsilon}{\partial T} = -\frac{1}{T^2 s''(\epsilon)\ }.
\end{equation}
The specific heat scales as $N^0$ in the hadronic regime and $N^2$ in the plasma phase.
Thus the ratio $r(T) \equiv c_V/\epsilon$ is order unity in either regime and remains finite as $N \rightarrow \infty$  except at putative points where $c_V$ diverges as at a 2nd order phase transition.  More generally $c_V$ is analytic except at phase transitions; thus $r(t)$ will be analytic at large $N$ except at phase transition points. Thus Assumption \ref{nonan} implies that in the large $N$ limit, $r(T)$ is finite and analytic everywhere except at $T_c$. 

The behavior  of $s(\epsilon)$ for homogeneous matter at large but finite $N$ in a intermediate regime between $\epsilon \sim N^0$ and $\epsilon \sim N^2$ must interpolate between the two regimes.  The precise way it does so is not determined from general scaling considerations.   This regime is of comparatively little interest in any case, since in this regime the system is locally unstable.

\begin{figure}[t]
  \begin{center} 
    \includegraphics[width=9cm]{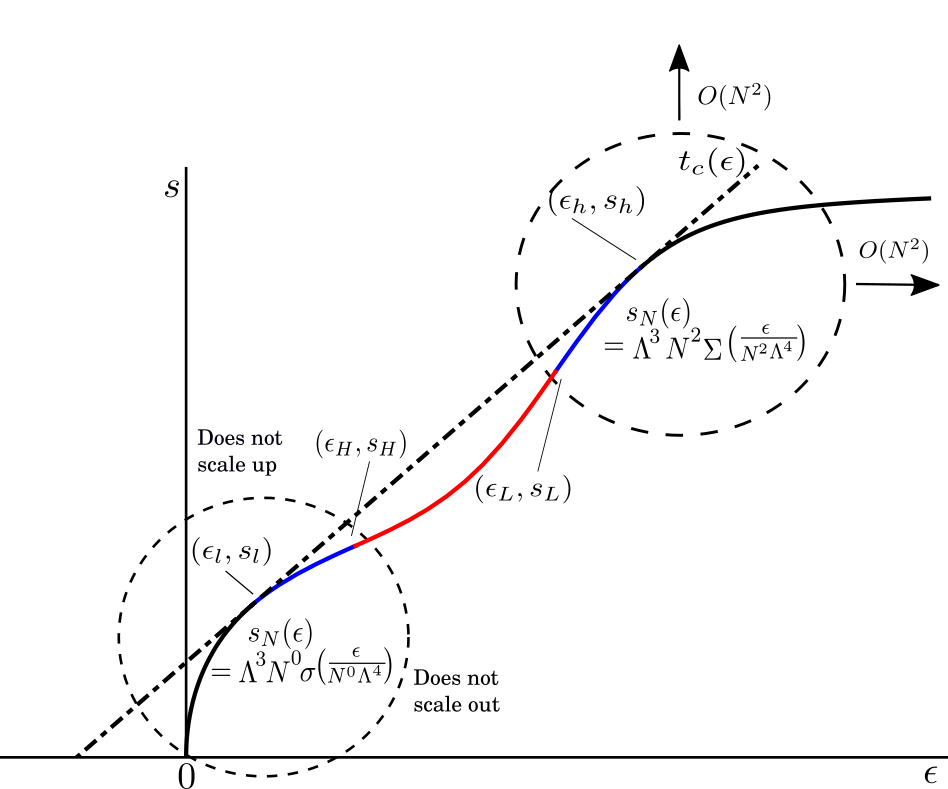}
    \caption{N scaling of entropy-energy density curve form homogenous matter.\label{fig:mc-LargeN} } 
  \end{center}
\end{figure}

Fig.~\ref{fig:mc-LargeN} summarizes the scaling of thermodynamic quantities in the two regimes and sketches the mixed phase, $t_c(\epsilon)$. The transition temperature and pressure are respectively denoted $T_c$ and $P_c$. The scalings given in Eq.~(\ref{sscale}) imply that  $T_c$ and  $P_c$ are independent of $N$ for large $N$.
 
These scalings imply a remarkable cancellation in the plasma phase.  Generically the plasma phase has a pressure of order $N^2$.  For example, at asymptically large energy densities, the pressure is given by $P=\epsilon/3 \sim N^2$.  However, at the phase transition point which is denoted  $\epsilon_h$ in Fig.~\ref{fig:mc-LargeN} a very large cancellation must take place:
\begin{equation}
\underbrace{P_c}_{\sim N^0}  = \underbrace{T_c}_{\sim N^0} \underbrace{s_{N}(\epsilon_h)}_{\sim N^2} - \underbrace{\epsilon_h}_{\sim N^2}\text.
\label{cancel}
\end{equation}
For $P_c$ to be of order $N^0$, the two terms of Eq.~(\ref{cancel}) must exactly cancel at leading order. The effects of this cancellation are central to Sec.~\ref{gluon}, where we show the existence of a strongly coupled plasma, and demonstrate that at large $N$ the supercooled plasma phase has negative absolute pressure.

Before we turn to Sec.~\ref{gluon}, it is important to note that this cancellation depends on the first-order transition occuring at a temperature for which the hadronic phase has an energy density of $N^0$. This is a consequence of Assumption~\ref{nonan}, which implies that in the large-$N$ limit, $r(T) \equiv c_V / \epsilon$ is finite and analytic throughout the hadronic phase. 
From the definition of  $r(T)$ it follows that the energy density at two different temperatures $T_a$ and $T_b$ are related by
\begin{equation}
\epsilon (T_b) = \epsilon (T_a) \exp\left( \int_{T_a}^{T_b} d T \, r(T)\right )\text.
\end{equation}
If the hadronic phase contained regimes with both $\epsilon \sim N^0$ and $\epsilon \sim N^2$, then taking $T_b$ to be in the latter and $T_a$ to be in the former, we would find that $r(T) \sim N^2$ in between, in contradiction to Assumption~\ref{nonan}. Therefore, if there were a hadronic regime with energy density of order $N^2$, at large $N$, it would need to be separated from the regime with energy density of order unity by a phase transition; such a putative phase transition would need to be between two distinct hadronic phases and would be in addition to the first-order phase transition between a hadronic and plasma phase.

Assumption \ref{Hagd} --- that the mesons and glueballs satisfy a Hagedorn spectrum --- is fully consistent with this behavior. 
Using the type of analysis discussed in Sec.~\ref{hagedorn} it is easy to see that  with a Hagedorn spectrum at large $N$, $\epsilon \sim N^0$ whenever $(T-T_H) \sim N^0$ and that $r(T)$ is nonanalytic at $T=T_H$\footnote{In particular, $\lim_{T \rightarrow T_H}\frac{d^n r(T)}{d T^n}$ diverges whenever $n \ge 9/2-d$, where $d$ specifies the  power law in the subexponential prefactor (as given in Eq.~(\ref{hage}).).  For $7/2< d \le 9/2$, $r$ itself diverges as $T \rightarrow T_H$}.   The assumption of a single first-order transition with a nonvanishing superheated phase implies that $T_c < T_H$.


\section{Plasma Phase}\label{gluon}
In this section we discuss the behavior of the low-temperature end of the plasma phase of large $N$ QCD, including the supercooled regime.

\subsection{Achieving negative absolute pressure}

In this subsection, we show that if the assumption that a generic first-order transition persists in the large $N$ limit with the latent heat growing with $N^2$, a supercooled plasma with negative absolute pressure --- a pressure less than the vacuum pressure --- must exist.   

Such a situation is quite unusual.  Consider the following {\it Gedankenexperiment}: suppose one had a rigid cylinder with a movable piston that was capable of containing large-$N$ QCD matter in the plasma phase.  The piston starts locked so the system has fixed volume and the system starts in the stable plasma phase. The cylinder is then brought into thermal contact with a heat bath whose temperature is slowly lowered so that the plasma in the cylinder equilibrates in the metastable phase.  At this point, the cylinder is thermally insulated so that it can no longer exchange energy with the outside and the system is isolated from other matter and is sitting in the vacuum.   Next, the piston is allowed to move freely. Remarkably, instead of pushing outward into the vacuum, it sucks inward.  This behavior is counterintuitive from the viewpoint of the kinetic theory of a gas, where particles in a box are hitting the wall of a chamber and transferring momentum to it when they bounce back; this necessarily results in a positive pressure. The breakdown of such intuition would indicate that the system is not describable, even qualitatively, as a plasma or gas of weakly interacting particles or quasiparticles. Issues regarding a possibly vanishing pressure in the plasma phase have been discussed in a more phenomenological context in~\cite{Kapusta:1979fh,Kapusta:1981ay,Kapusta:1981ue,Kapusta:1982qd}.

Moreover, a negative absolute pressure runs counter to intuition gleaned from stable phases.   One does not come across stable phases with negative absolute pressure for stable phases for systems with zero chemical potentials.  Indeed, it is easy to see that a negative absolute pressure is impossible for such systems provided that they also satisfy the condition that only positive temperatures are possible. This follows from the requirement of global stability, which implies that $s'(\epsilon_1) \ge s'(\epsilon_2)$ where $\epsilon_1< \epsilon_2$. The temperature is thus everywhere a nondecreasing function of $\epsilon$ (provided $T$ is nonnegative).   This in turn implies that for $\epsilon>0$,
\begin{equation}
s(\epsilon)=\int_0^{\epsilon} {\rm d} \tilde\epsilon \, s'(\tilde{\epsilon}) > s'(\epsilon) \epsilon\text,
\end{equation}
and therefore the pressure $P = s/s' - \epsilon$ is positive.

This argument holds only for the curve $s(\epsilon)$ describing globally stable matter. It does not apply to a metastable supercooled phase.  The supercooled phase does not lie on the purely concave downward curve for the stable phase. Rather, as seen in Fig.~\ref{fig:mc-LargeN} it lies along the curve for homogenous phases which includes regions that are concave upward as well as regions that concave downward. Thus, systems need not have positive pressure relative to the vacuum in the supercooled phases. 

The scaling relations of Eq.~(\ref{sscale}) along with the cancellation seen in Eq.~(\ref{cancel}) imply that not only is negative pressure possible for the supercooled phase of large $N$ QCD, it is necessary provided that  Assumption \ref{nonan} holds. We assume the existence of a supercooled plasma phase over a nonzero range of temperature in the large-$N$ limit in the following discussion. At the transition temperature $T_c$, cancellations ensure that the pressure is order unity, so that $T_c s_h = \epsilon_h + O(N^0)$ ($\epsilon_h, s_h$ are as defined in Fig.~\ref{fig:mc-generic}).  Recall that for zero chemical potential the Helmholz free energy is minus the pressure.  Thus,  $\frac{\partial P}{\partial T} = s$ with $ s \sim N^2$ in the supercooled phase.  Given these conditions, consider what happens as we lower the temperature from $T_c$ to some temperature $T_{sc}$ in the supercooled phase, an order-unity distance below $T_c$.  The pressure decreases by an amount of order $s T \sim N^2$. However, the pressure was only of order unity at $T_c$, and therefore must be negative at $T_{sc}$. Algebraically:
 \begin{align}
P(T_{\rm sc}) &=  P(T_c) - \int_{T_{\rm sc}}^{T_c} {d} T  \frac{d P}{d T}
\nonumber\\
&< \underbrace{P(T_c)}_{\sim N^0} - \underbrace{(T_c - T_{sc})}_{\sim N^0} \underbrace{s(T_{sc})}_{\sim N^2}
\end{align}
where the inequality follows from the fact that $s_N(\epsilon)$ is positive and monotonically increasing  with $T$. This argument is quite simple and the conclusion is striking: given the assumptions outlined in the introduction, at large $N$, the supercooled region has negative absolute pressure. An analogous discussion of negative absolute pressure to the one above was given in~\cite{Aharony:2005bm} in the context of their analysis of ``plasma-balls''.

A key part of Assumption~\ref{nonan} is that a supercooled metastable phase persists in the large-$N$ limit over a nonzero range of temperatures. It is plausible that this is not the case --- that in the large-$N$ limit, the supercooled phase vanishes.
Based on the analysis of supercooled metastable phase above, one of the following must be true if a first order: As $N$ gets large either the medium achieves absolute negative pressure in a supercooled regime, or  no supercooled regime exists  in the large-$N$ limit.

\subsection{A strongly interacting plasma}

Next we consider the implications of $N$ scaling for the  existence of strongly coupled plasma at large $N$. In QCD with $N=3$ and physical quark masses,  the hadronic regime goes over to the quark-gluon regime via a continuous crossover as the temperature increases; there is no true phase transition. In the vicinity of the crossover, there is strong empirical evidence that the system is strongly coupled. For example, the ratio of viscosity to entropy density, $\eta/s$, extracted from hydrodynamic simulations of heavy-ion collisions is small,  $\eta/s \sim 1/4\pi$~\cite{PhysRevLett.106.192301} --- which is an indication that the system, whatever it is, must be strongly coupled.  Conventionally, this medium has come to be called a strongly interacting quark-gluon plasma (sQGP).  

While the evidence that the system is strongly coupled is compelling, its description as a quark-gluon plasma might be regarded as less so.  The principal logic behind such a description is that the energy density is too high for the system to be described as a gas of clearly discernible hadrons.  While this is true, it is also true the system is not in a regime where it can be described as behaving as a plasma of clearly discernible quarks and gluons (as one has in  a weakly coupled plasma).   Thus the description as an sQGP as opposed to a strongly interacting hadronic gas could be thought of as merely a convention.  

This raises an interesting issue of principle: does there exist a gauge theory that shares with $N=3$ QCD an unambiguously hadronic regime and an unambiguous plasma, but also has the feature there exists a regime that is both unambiguously strongly coupled and unambiguously a plasma?  If the answer to this question is in the affirmative, it gives at least some justification for the conventional description of an sQGP in QCD for $N=3$ where the situation is more ambiguous.

In this section we will see that the large-$N$ limit of QCD is precisely such a theory. The first-order transition cleanly separates two phases with two different $N$-scalings, and thus defines unambiguous hadronic and plasma regimes. In what follows, we will show that as the energy density approaches $\epsilon_h$ from above, the system becomes arbitrarily strongly coupled.  It may not be obvious how to measure the strength of the interaction for this system. With present numerical and theoretical methods, there is no practical way to determine $\eta/s$ for a large-$N$ QCD system.  One could imagine a computation of $\eta/s$ from first-principles lattice QCD simulations; however such numerical evaluation of the shear viscosity has technical issues stemming both from the largeness of $N$~\cite{gattringer2009quantum} and the real-time nature of the observable~\cite{Alexandru_2016,Cohen_2015}.

So what are the other useful observables to distinguish strongly coupled from weakly coupled systems?   For massless constituents such as gluons, a useful measure of the strength of the interaction between constituents in a medium is the ratio $\epsilon/3P$. For a non-interacting system of massless particles, this ratio is $1$; it deviates from unity for a system of massive particles or for a strongly interacting medium.  Thus the condition $\epsilon/3P \gg 1$, in a plasma might be taken as a signal for a strongly coupled plasma. 
One might worry that the quarks in a quark-gluon plasma are not massless.  However, in the plasma phase, their contributions are  suppressed by relative order $1/N$.

It is not clear {\it a priori} how large $\epsilon/3P$ should be for the system to be identified as strongly coupled.  However, at large $N$, the ratio can be made arbitrarily large, even as the system remains in the plasma phase.  This occurs in the double limit $N \rightarrow \infty$, $\epsilon \rightarrow \epsilon_h^{(+)}$, where the superscript $(+)$ indicates that the limit is taken from above to ensure the system is in the plasma phase.  The unbounded growth happens in this limit since in the plasma phase, the energy density is of order $N^2$ throughout and this extends down to $\epsilon_h$. However, the pressure is of order $N^0$ at $\epsilon_h$.   Thus ratio $\epsilon/3P$ diverges in the large-$N$ limit, and so the plasma can be made arbitrily strongly coupled. It is worth noting that if the plasma supercools at large-$N$, as discussed above, the ratio $\epsilon/3P$ wraps around infinity to become negative, remaining strongly coupled.

In summary we have demonstrated that at large $N$, there exists a regime in which the system is unambiguously in the plasma phase and strongly coupled.  In light of the issue of principle set out at the beginning of this subsection, it is worth noting that while a strongly coupled plasma regime exists at large $N$, a strongly coupled hadronic regime does not.   This gives at some modest support to the notion that the strongly interacting medium seen empirically in QCD with $N=3$ might be identified as being an sQGP as opposed to strongly interacting hadronic gas.


\section{Superheated Hadronic Phase and Beyond}\label{hadron}
Now we move to considering the hadronic phase. In the large-$N$ limit this phase consists of a non-interacting gas of hadrons. The analysis in this section is based on Assumption \ref{Hagd} from the introduction: we assume an exponential Hagedorn spectrum with $d > 7/2$ for a subexponential prefactor $m^{-d}$. The motivation for this assumption is discussed below.

In Subsection~\ref{hagedorn} key thermodynamic quantities such as $s$ and $\epsilon$ are discussed in terms of the Hagedorn spectrum.  A key aspect of this analysis is the assumption  $d > 7/2$.  It  implies that at large $N$, as the temperature approaches the Hagedorn temperature $T_H$, the energy density remains finite and independent of $N$.  We will denote the limiting energy density as $T \rightarrow T_H$ by $\epsilon_H$.  Subsection~\ref{unstable}  considers the extension of $s(\epsilon)$ curve into the locally unstable regime with $\epsilon > \epsilon_H$ in the $N\rightarrow\infty$ limit.

All of the analysis in this section is aimed at the hadronic regime.  The properties of this regime are independent of $N$ in the large $N$ limit.  Accordingly, in all of the analysis, we take the large $N$ limit at the outset, unless otherwise specified.  Energy densities play a central role in the analysis; in this section we will assume that the energy densities under consideration do not  grow with $N$ as $N$ gets large --- they scale as $N^0$.

\subsection{Hagedorn spectrum}\label{hagedorn}

In the $N\rightarrow \infty$ limit, the maximum temperature of the locally stable hadronic phase is the Hagedorn temperature, $T_H$. It is necessary that $T_c \le T_H$, but whether the inequality is strict is unknown from first principles. If the inequality is strict, then there exists a metastable superheated hadronic phase; if not, then this metastable phase (if it exists at all) must disappear in the large-$N$ limit.  In the analysis here, Assumption \ref{nonan} requires a generic first-order  phase transition that allows superheating and so we take the inequality to be strict.

It has long been believed that large-$N$ QCD has a Hagedorn spectrum~\cite{Thorn:1980iv}.  This belief is partly motivated by the fact that that highly excited states at large $N$ appear to act like excitations of long flux tubes, which may be regarded as string-like; string theories have Hagedorn spectra~\cite{Zwiebach:2004tj}.  However, there  is a compelling argument that  large-$N$ QCD has a Hagedorn spectrum that does not explicitly assume stringy dynamics; rather it is  based on commonly accepted properties of QCD correlation functions and the fact that the number of local operators of fixed mass dimension grows exponentially with the mass dimension~\cite{Cohen:2009wq}. 
Supported by arguments above, it is believed that the spectrum of mesons and glueballs  at large $N$ is a Hagedorn spectrum ~\cite{Hagedorn:1965st}, with the number of hadrons $N^{\rm had}(m)$ of mass less than $m$ governed at asymptotically large masses by
\begin{equation}
\label{hage}
\begin{split}
N^{\rm had}(m)& = C {m^{-d}}{T_H^d} \exp(m/T_H) + \Delta(m) \\
& \text{with} \;  \frac{ \Delta(m)}{N^{\rm had}(m)}  =  {\cal O}(1);
\end{split} \end{equation} 
the constant C is a dimensionless numerical factor, and $\Delta(m)$ is a correction term that, at large $m$ is asymptotically smaller than $N^{\rm had}(m)$ . Note that $N^{\rm had}(m)$ increases by discrete steps of unity as $m$ increases. While the leading term is smooth, the discrete behavior is encoded in $\Delta (m)$. Given the current state of the art, the power law prefactor specified $m^{-d}$ in the spectrum Eq.~(\ref{hage}) cannot be determined from first principles theoretically nor can it be determined reliably from fits to the currently available spectrum~\cite{Cohen:2011cr,PhysRevC.92.055206}. Although Hagedorn originally proposed $d=5/2$~\cite{Hagedorn:1965st}, there are very good reasons to believe that  $d=4$.  A simple bosonic string theory that has $d=4$~\cite{Zwiebach:2004tj}; such a string theory is natural at large $N$ if flux tube dynamics dominate~\cite{Aharony:2013ipa}. Here we assume that $d > 7/2$ so that energy density and entropy density are finite up to $T_H$ as is discussed below.

The value of $d$ plays a nontrivial role in the large-$N$ QCD thermodynamics~\cite{Thorn:1980iv,Cohen:2006qd}.  The Hagedorn spectrum creates divergences in various thermodynamic quantities due to the exponential growth in the number of hadrons with mass, but the nature of these divergences depends strongly on $d$. We begin by establishing that the entropy density and energy density diverge as $T$ approaches $T_H$ from below, only when $d \le 7/2$. As the divergences are related to the large-mass asymptotics of the Hagedorn spectrum, we are justified in working in the limit of $m \gg T$. For an ideal gas of one species in this limit, the energy and entropy density are given by
\begin{eqnarray}
\epsilon &=& \frac{1}{(2\pi)^{3/2}}e^{-m/T}m^{5/2}T^{3/2} 
\left (1+ {\cal O}\left( T/m\right ) \right )\text{,}  \label{e}\\
s &=& \frac{1}{(2\pi)^{3/2}} e^{-m/T}m^{5/2} T^{1/2}
\left (1+ {\cal O}\left( T/m\right ) \right )
\text.\label{s}
\end{eqnarray} 

To understand the behavior of $\epsilon$ (and equivalently, $s$) for $T$ near $T_H$ --- which we refer to as the Hagedorn point --- it helps to divide the energy density into contributions from low and high mass hadrons, split arbitrarily at some large mass $m_0$:
\begin{equation}
\epsilon(T) =\epsilon^{\rm low}(T;m_0) +\epsilon^{\rm high}(T;m_0)\text.
\end{equation}
The mass $m_0$ is chosen to be much larger than $T_H$ --- large enough that the asymptotic form of the Hagedorn spectrum is accurate.

After fixing $m_0$, $\epsilon^{\rm low} (T;m_0)$ is an analytic function of $T$, as it is the sum of the contributions from a finite number of species. The form of the Hagedorn spectrum does not determine $\epsilon^{\rm low}$, but neither is $\epsilon^{\rm low}$ relevant to understanding divergences at the Hagedorn point.

From Eqs.~(\ref{hage}) and (\ref{e}), the contribution from the large-mass portion of the Hagedorn spectrum is given by
\begin{align}
\epsilon^{\rm high} &(T; m_0) = \int_{m_0}^{\infty} dm\,\,\left( \frac{d N^{\rm had}(m)}{d m} \right )\,\epsilon(m,T) \nonumber\\
&\approx \frac{C}{(2\pi)^{3/2}} {T^{3/2}}{T_H^{d-1}}   \int_{m_0}^{\infty} \!\d m \; m^{5/2-d} e^{-m\left(\frac 1 T - \frac 1 {T_H}\right)}\text.
 \label{eT}
\end{align}
Two approximations have been made, both becoming arbitrarily accurate as $m_0$ is increased. First, we have assumed the large-mass limit in Eq.~(\ref{e}). Second, in neglecting subleading terms in $m^{-1}$, we are permitted to replace the discrete sum over masses by an integral and treat the Hagedorn spectrum as if it were continuous.

The integral in Eq.~(\ref{eT}) converges for any $T<T_H$ and diverges for any $T>T_H$.  Because $\epsilon^{\rm high}$ contains the entire nonanalytic part of $\epsilon$, we see that for any fixed $m_0$, $\epsilon(T, m_0)$  is a nonanalytic function of $T$ at $T=T_H$.  It is important, however, to recall that the behavior is only required to be nonanalytic in the large $N$ limit.  

It has long been known that 
the precise nature of the nonanyticity at $T = T_H$ depends on the value of $d$~\cite{Thorn:1980iv,Cohen:2006qd}. We see from Eq.~(\ref{eT}) that when $d \le 7/2$ the integral diverges and thus $\epsilon$ goes to infinity as $T_H$ is approached from below.  However for $d >7/2$, as is assumed here, the integral converges even at $T = T_H$: a finite energy density is attained in the limit $T \rightarrow T_H$.  We denote this this value $\epsilon_H$.   

Although the integral in Eq.~(\ref{eT}) converges when $d > 7/2$, the nonanalyticity is manifested in the divergent derivatives of $\epsilon$ with respect to $T$ as $T \rightarrow  T_H$:
\begin{equation}
\frac{d^n \epsilon(T)}{d T^n} \sim \frac{1}{(T-T_H)^{n+d-7/2}} \; \; {\rm for} \; \; n+d-7/2 > 0\text.
\end{equation}

The situation for the entropy density is essentially identical. Decomposing $s(t) = s^{\rm low}(T;m_0) +s^{\rm high}(T;m_0)$, a quick calculation reveals that
\begin{equation}
s^{\rm high} (T) \approx  \frac{\epsilon^{high}(T)}{T}\text.
\label{sT}
\end{equation}
Clearly $s(T)$ is also nonanlytic at $T_H$ and also has a finite limit as $T \rightarrow T_H$ from below when $d > 7/2$.

For the remainder of this section we will employ Assumption \ref{Hagd} and take $d > 7/2$. With this assumption, both the energy density and entropy density are finite at large $N$ in the entire locally stable hadronic phase, but are ill-defined at any $T > T_H$.    

We note that if $d=4$ (as expected from  string theory) --- or more generally for any $d$ satisfying $9/2 \ge d >  7/2$ --- while the energy density remains finite at large $N$ as $T \rightarrow T_H$, the specific heat $c_V$ diverges at the Hagedorn point and the speed of sound, $c_s =\sqrt{\frac{\partial P}{\partial \epsilon}} = \sqrt{\frac{s}{c_V}}$ goes to zero.  

An interesting question  arises when the energy density of a system composed of hadrons exceeds $\epsilon_H$.  We describe such a system as having an energy density beyond the Hagedorn point.

\subsection{Beyond the Hagedorn point}\label{unstable}

At large $N$, the maximum temperature in the hadronic phase is $T_H$.  Moreover, if $d > 7/2$ as we are assuming here, the energy density remains finite, limiting to $\epsilon_H$, as $T \rightarrow T_H$ from below.  

This creates a somewhat paradoxical situation. Consider a very large box in the metastable superheated phase at a temperature just below $T_H$, so that the energy density is just below $\epsilon_H$.  Now suppose that a number of hadrons with a total energy density of order unity and sufficient to raise the energy density above  $\epsilon_H$ are carefully injected into the system.  Given that $T_H$ is the maximum temperature, the temperature cannot increase while the system remains in the hadronic phase.  The paradox is that at large $N$, hadrons are very weakly interacting, so that there is nothing to keep us from  continuing to introduce  hadrons into a large box until their energy density exceeds $\epsilon_H$, with the system still being composed of clearly identifiable hadrons.   It is difficult to reconcile this fact with $T_H$ being the upper bound for temperature. This is at least suggestive of a breakdown of the canonical ensemble (as noted in~\cite{Carlitz:1972uf}), so we will proceed with an analysis in the microcanonical ensemble.

\begin{figure}[t]
  \begin{center} 
    \includegraphics[width=9cm]{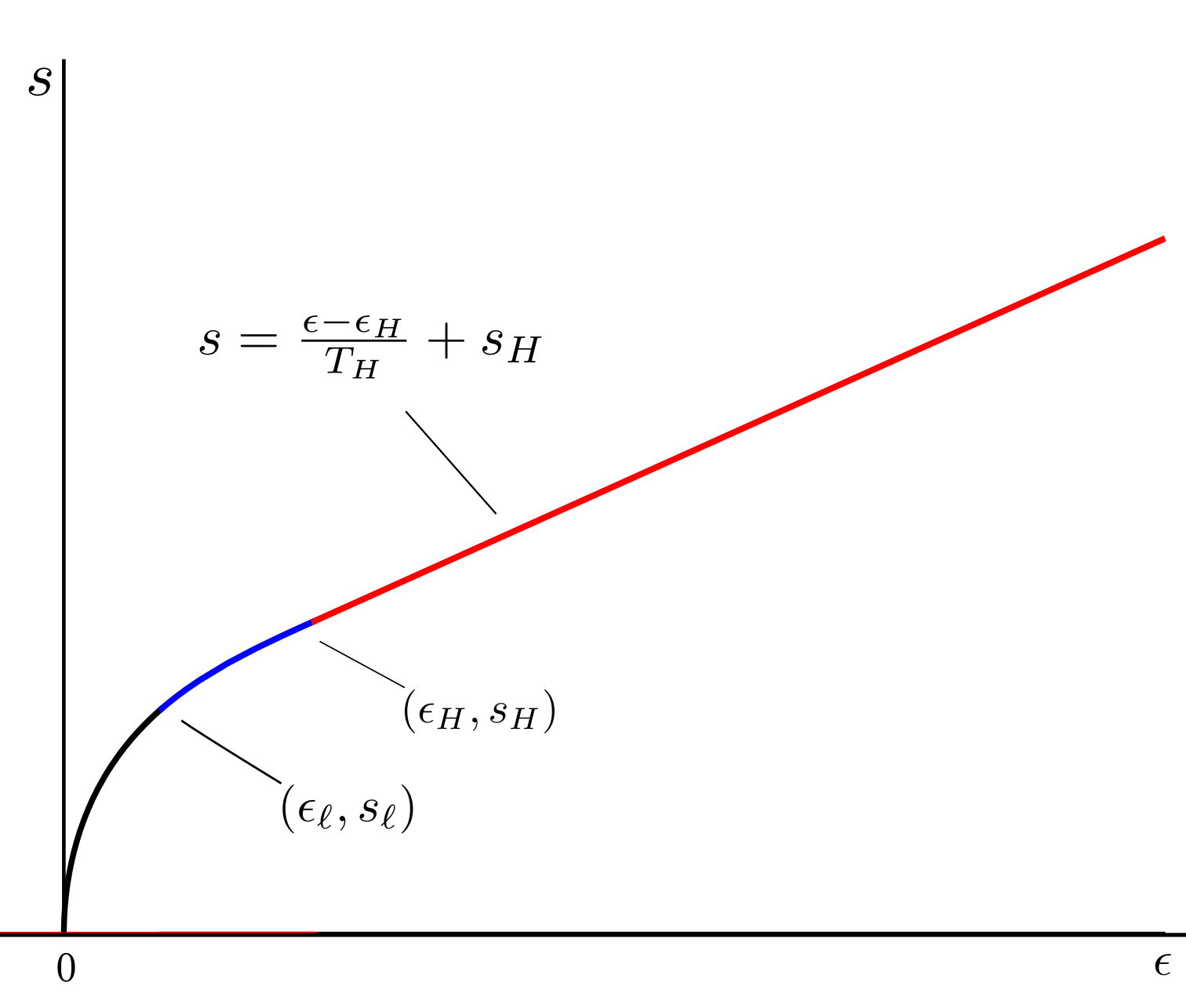}
    \caption{The behavior of $s(\epsilon)$ at large $N$, including the region $\epsilon >\epsilon_H$. \label{Beyond}}
    \end{center} 
\end{figure}

A natural way to try to reconcile these is to focus on the behavior of the curve $s(\epsilon)$  for homogenous matter, into the region with $\epsilon > \epsilon_H$; this is the regime that is locally unstable for a generic first-order transition.  We will show that in the limit $N \rightarrow\infty$, the curve $s(\epsilon)$ extends past $\epsilon = \epsilon_H$ linearly, with the temperature (given by the inverse slope) constantly equal to $T_H$ .  This is illustrated in Fig.~\ref{Beyond}.

One consequence of this behavior is that there is a discontinuity in the third derivative $\frac{\partial^3 s}{\partial\epsilon^3}$ at the Hagedorn point.   Such behavior is typically seen  at a second-order phase transition.  Beyond $\epsilon_H$, as $\epsilon$  increases the temperature remains fixed, as typically happens for a mixed phase after a first-order transition\footnote{Such behavior seen for $d=4$ is expected if large-$N$ QCD becomes a string theory, and more generally holds if $9/2 \ge d >7/2$.  Note that if $d > 9/2$, there is also a discontinuity of the second derivative.}.  The behavior containing features of both first- and second-order transitions is quite unusual.  It may be a hint that at large $N$, the systems does not in fact behave like a generic first-order transition, invalidating Assumption~\ref{nonan}.  In the remainder of this section, however, we will assume that Assumption \ref{nonan} remains correct and that higher-order $1/N$ effects in the region beyond $\epsilon_H$ reconcile this behavior with a  generic first-order transition.

Let us see how this behavior comes about.  To do so we must calculate $s(\epsilon)$, in the $N\rightarrow\infty$ limit, in the region $\epsilon > \epsilon_H$.  While our interest is in $s(\epsilon)$, a quantity which arises naturally in a microcanonical description, we will exploit the equivalence in the thermodynamic limit of large volumes between predictions using canonical and microcanonical ensembles.  As will be seen below this equivalence breaks down when   $\epsilon > \epsilon_H$, but the way it breaks down will allow us to determine the behavior for  $\epsilon > \epsilon_H$.

 Up to the Hagedorn point,  where $\epsilon=\epsilon_H$, one may work with the canonical ensemble to calculate both $s$ and $\epsilon$ as a function of $T$;  the microcanonical curve $s(\epsilon)$ is obtained parametrically by varying $T$. This method works up to the Hagedorn point.  However, if one tries to go beyond $\epsilon_H$ by increasing the temperature, the canonical expressions diverge. 
 
To proceed beyond the Hagedorn point, we will first impose a constraint on the class of microstates we consider.  This is a legitimate procedure in microcanonical physics.  If one computes $s^{\rm const}(\epsilon)$ --- the logarithm of the number of microstates subject to the constraint, one knows that by construction, the unconstrained entropy density $s(\epsilon)$, which is calculated with a superset of those constrained states, must satisfy $s(\epsilon) > s^{\rm const}(\epsilon)$.   The constraint we impose is that only configurations in which all of the hadrons in the system have masses that are less than  some large value, $m_{\rm max}$, are considered.  This constraint might be envisioned either as arising in some possible physical realization (at least in principle\footnote{One could envision a thought experiment in which some active device identifies and removes all hadrons with masses above $\mmax$ that are created by those rare interactions that equilibrate the system and replaces them by two or more lighter hadrons carrying same energy.}) or simply as a mathematical device to facilitate counting.   In practice, it is easy to see how this works: the constraint cuts off at $\mmax$ the contributions of the high mass hadrons --- those responsible for the divergences --- and thereby renders finite and analytic $\epsilon$ and $s$. Ultimately we consider the behavior of the system when we remove the constraint by letting $\mmax$ become arbitrarily large.

When the constraint is imposed, the problem becomes one of a relativistic ideal Bose gas containing a finite --- albeit very large --- number of species.  The fact the number of (noninteracting) species is finite ensures the microcanonical and canonical descriptions agree. This holds whether or not $\epsilon >\epsilon_H$. Note moreover that with a finite number of species of free particles, there is no possibility of a phase transition so that subject to the constraint,  $s(\epsilon)$ is an analytic function; nonanalyticities can arise only when the $\mmax \rightarrow \infty$ limit is taken.  Our goal is to determine the  $\mmax \rightarrow \infty$ limit of $s(\epsilon; \mmax)$, the entropy density as a function of energy density subject to the constraint that all hadrons have masses below $\mmax$ in the region $\epsilon>\epsilon_H$. To do so we will first determine the limit as  $\mmax\rightarrow\infty$ of $s''( \epsilon)^2$, and then integrate twice to find $s(\epsilon; \mmax)$.

The most straightforward way to proceed is to compute $s''(\epsilon)$ for the hadronic regime with  $\epsilon >\epsilon_H$  and asymptotically high values of $\mmax$. The calculation is somewhat long  but is essentially straightforward.  One obtains:
\begin{equation}\label{princip}
\begin{split}
&s''(\epsilon)= \frac{1}{(\epsilon_H-\epsilon) m_{\rm max}}  \times\\ &  \left ( 1 + \frac{1}{\left(d-\frac72\right) \log\frac{m_{\rm max}}{T_H}}+ {\cal O}\left( \frac{1}{\left(\log^2\frac{m_{\rm max}}{T_H}\right)  } \right ) \right ).
\end{split}\end{equation}
Clearly as one takes $\mmax$ to infinity, $s''(\epsilon)$ goes to zero, and $s'(\epsilon) =1/T$ is a constant.   The value of that constant, is its value at the low end of the regime\footnote{There is a potential subtlety here.  While $s(\epsilon)$ is an analytic function for an  finite value of $\mmax$, it need not be  in the $\mmax \rightarrow \infty$ limit. 
In fact, the regime of $m_{\rm max}$ for which Eq.~(\ref{princip}) holds, depends on the value of $\epsilon-\epsilon_H$; the regime is pushed off to infinity as  $\epsilon-\epsilon_H$ approaches zero.  The underlying reason for this is easy to understand: $s(\epsilon)$ is non-analytic at $\epsilon=\epsilon_H$ when  $m_{\rm max}$ is infinite but is analytic for any finite  value of   $m_{\rm max}$.  Thus, the limiting behavior when $m_{\rm max}$  gets large and $\epsilon-\epsilon_H$ get small depends on the order in which limits are taken.  However, our goal is to study the $\mmax \rightarrow \infty$ limit of the system, and in that limit $s''(\epsilon)=0$ in the entire $\epsilon > \epsilon_H$ regime extending  all the way down to  $\epsilon_H$.
 Ultimately, it is straightforward to show that $ \lim_{\epsilon \rightarrow \epsilon_H^+}  \, \left ( \lim_{\mmax \rightarrow \infty }\,  s'(\epsilon) \right )=1/T_H$. }, $1/T_H$, and the form seen in  Fig.~\ref{Beyond} emerges.


Equation (\ref{princip}) depended on a somewhat involved calculation.  However, one can intuitively understand the form of the $s(\epsilon)$ beyond the Hagedorn point by a very simple, if indirect argument.  Recall that If we fix $\mmax$ to be very large but finite,  there is no possibility of a phase transition. Therefore, $T(\epsilon;\mmax)$ must be a monotonically increasing function of $\epsilon$. In particular, we obtain the lower bound $T(\epsilon;\mmax) > T_H$ for $\epsilon > \epsilon_H$. The limit as we lift the cutoff, therefore, is similarly bounded below, $T(\epsilon) \ge T_H$, with equality now permitted.

Since the temperature in the canonical description, at $\mmax\rightarrow\infty$, cannot be permitted to exceed $T_H$, this suggests that $T(\epsilon) = T_H$ for all $\epsilon$ above $\epsilon_H$ (recalling that in this section we only consider energy densities that scales as $N^0$). Thus $s(\epsilon)$ is linear in that regime, with a slope determined by the Hagedorn temperature.

In fact, we can be more rigorous in determining the upper bound on $T(\epsilon)$, without invoking Eq.~(\ref{princip}). Fixing a temperature $T' > T_H$, observe that $\epsilon(T'; \mmax)$ can be made arbitrarily large by increasing $\mmax$, due to the Hagedorn divergence. This implies that, for any desired $\epsilon'$, and any $T' > T_H$ (no matter how close to the Hagedorn temperature), we can find some finite $\mmax$ for which $\epsilon(T';\mmax) > \epsilon'$, and therefore $T(\epsilon';\mmax) < T'$. As the limit $\mmax\rightarrow\infty$ is taken, $T(\epsilon;\mmax)$ is thus sandwiched between a lower bound of $T_H$ and an upper bound which asymptotically approaches $T_H$ from above, and we are forced to conclude that $T(\epsilon) = T_H$ for all $\epsilon > \epsilon_H$ (in the order-one regime).

This resolves our puzzle:  one can introduce energy into a hadronic system so that $\epsilon > \epsilon_H$ without pushing the temperature beyond $T_H$; $T$ hits $T_H$ and stays there.  But this creates a new paradox. Clearly, the microcanonical and canonical ensembles differ in this regime, since knowledge that $T=T_H$ is not sufficient to determine the energy density.  However, for any given species of hadron,  the canonical description should be valid and thus  determine the energy density associated with that species.  If one sums these together one necessarily gets $\epsilon=\epsilon_H$. The paradox is that there is more energy in the system than this.

The resolution to this paradox is that the standard thermodynamic behavior of energy density becoming an intensive quantity in the limit of large volumes breaks down.  Since it is only in the thermodynamic limit that the canonical and microcanonical descriptions agree, there is no incompatibility. 

To see why the standard thermodynamic limit breaks down, let us again introduce the constraint that only hadrons with a mass less than $\mmax$ contribute. With this cutoff, we can determine which hadrons dominate $\epsilon - \epsilon_H$ when the system is beyond the Hagedorn point.  By taking the cutoff $\mmax$ to be arbitrarily large, we can take the temperature $T$ at the desired energy density to be arbitrarily close to $T_H$. The difference in energy densities is then determined by the specific heat.  The specific heat diverges as $T\rightarrow T_H$ from below in the absence of the constraint, but it remains finite when the constraint is imposed.  Thus, the divergent behavior in the specific heat is dominated by contributions from large-mass hadrons. Therefore, it is highly plausible  that with a cutoff imposed, the excess energy density would be dominated by hadrons at the scale of the cutoff.
When the constraint is released, and $\mmax$ is taken to infinity, these contributions are pushed to arbitrarily high masses. In effect, all of the light hadrons equilibrate normally as an ideal gas at $T_H$, and all of the extra energy is pushed to arbitrarily high masses.

This intuitive argument can be made  precise.  In the hadron regime at large $N$, the energy density of the system  is constructed from the contributions of  the various species of noninteracting hadrons.   If we define $\epsilon_j(\epsilon,\mmax)$ as the contribution of the $j^{\rm th}$ species of hadron to the energy density, when the total energy density is $\epsilon$ and a constraint of including only hadrons with a mass below $\mmax$ is imposed.   When $\epsilon > \epsilon_H$, one can now define $\langle m \rangle$ to be the energy-weighted average mass of the hadrons contributing to the difference between $\epsilon$ and $\epsilon_H$ as
\begin{equation}
\langle m \rangle =\frac{ \sum_j m_j \left(\epsilon_j(\epsilon,\mmax)-\epsilon_j(\epsilon_H,\mmax)\right )}{ \sum_k \left(\epsilon_k(\epsilon,\mmax)-\epsilon_k(\epsilon_H,\mmax)\right )} \, ;
\end{equation}
 $m_j$  is the mass of the the $j^{\rm th}$ hadron.  A straightforward but somewhat involved calculation gives the asymptotic form at  large $\mmax$ of  $\langle m \rangle$ :
\begin{equation}
\langle m \rangle_{\rm asympt} = m_{\rm max} \left( 1- \frac{1}{(d-\frac72)\log\left(\frac{ m_{\rm max}}{T_H}\right)} \right ) \, . 
\label{mtyp}\end{equation} 
 As expected, when $\mmax$ gets large, so does mass of a typical hadron contributing to the energy density above $\epsilon_H$.  It is striking that the typical mass is not merely at the scale of $\mmax$, it is parametrically close to  $\mmax$ itself.

Now consider what happens if one drops the constraint of $\mmax$ and has a large volume $V$, into which energy $E$ in the form hadrons is injected such that $\frac E V >\epsilon_H$.  Note that even without  $\mmax$, the finiteness of the system imposes an upper bound of $E$ on  the possible masses.  Thus, one could    impose $\mmax = E$ without changing the physics.  If one assumes a standard thermodynamic limit with an intensive energy holds, one sees from that all of the additional energy between $E$ and $\epsilon_H V$ is contained in hadrons with typical masses very close to $E$ itself.  One has virtually all of the excess energy in a single hadron. But this is true regardless of the volume of the system, which is incompatible with the assumption of a standard thermodynamic limit; the assumption of a standard thermodynamic limit implies its own contradiction.
 
While the lack of a well-defined thermodynamic limit is a reasonable mathematical explanation, it is important to understand what is happening physically.  
The physical picture is actually quite simple;  after additional energy is injected into the system raising the average energy density above $\epsilon_H$, interaction effects (that are subleading in $1/N$) would continuously rearrange the energy so the lighter hadrons would move toward distributions compatible with the temperature of $T_H$, and  the additional energy would flow to higher and higher masses;  for a system with infinite volume this process  would never stop and the system would never fully equilibrate.   This thermodynamic system would appear to exemplify the maxum that ``there is always room at the top''.   For a finite volume system, this process would have to stop --- eventually,  the mass of the hadrons absorbing the excess energy would become comparable to the excess energy itself and the system can equilibrate.   However, for large volumes the equilibration time becomes long.  
 
Of course, the behavior of infinite system with energy flowing to ever higher masses cannot be realized in practice. For one thing, real all systems have finite volume.  Another obvious reason is that $N$ is finite for any conceivable physical system --- and rather small in QCD itself.    There are two principal effects associated with finite $N$.   The first effect is associated with the masses of the hadrons.  As the masses of the hadrons are pushed ever upwards, in this scenario, they will eventually reaches a point at which they can no longer be regraded as being of order unity in a $1/N$ expansion.  However, the notion of well-defined narrow mesons and glueballs with well-defined masses, which  is the basis of this analysis, is only valid for hadrons with masses of order unity.    Secondly,  we have used ideal gas expressions since hadron-hadron interactions are subleading in the $1/N$ approximation.   Presumably, these subleading effects restore a well-defined thermodynamic limit as volumes increase --- but when $N$ gets large it is approached very slowly.
 
These $1/N$ corrections are essential if Assumption \ref{nonan} of the introduction is to hold.  It assumes a single  generic first order phase transition.  This implies that in  the region where $\epsilon > \epsilon_H$,  $s''(\epsilon) >0 $.  However, at large $N$, $s''(\epsilon) =0$; instead of corresponding to a locally unstable system as expected in a generic first order transition, it is neutral, with neither positive or negative feedback from small amplitude fluctuations of large size. However, higher-order $1/N$ corrections can ensure  that it $s''(\epsilon) >0$, even if it is  small.  In that case one can have a generic first-order transition between a hadron regime with $\epsilon$ and $s$  are both of order $N^0$ and a plasma regime where they are both of order $N^2$.


\section{Discussion}\label{Dis}

To summarize the principal results of this paper:  Given the three apparently innocuous assumptions of the introduction, QCD (and pure gauge theory) in the large $N$ limit:
\begin{enumerate}
\item The system has negative absolute pressure --- a pressure below the theory's vacuum --- in the supercooled metastable plasma phase.
\item  There exists a regime in which the theory is both unambiguously in the plasma phase and unambiguously strongly coupled.
\item A well-defined thermodynamic limit does not exist in the hadronic regime, when the energy density exceeds $\epsilon_H$. 
\end{enumerate}

Of course, one possibility is that these assumptions are not innocuous.  For example, one might imagine that Assumption \ref{nonan} is not correct in a subtle way such that for every large value of $N$ a first order transition exists with finite temperature domain of supecooling for the plasma phase, but the size of this domain drops to zero as $N$ approaches infinity.  Were that to occur, the conclusion that a negative absolute pressure well occur in the supercooled phase can be evaded.  However, such a situation would be very interesting in its own right, since unlike in a typical first order transition, there would be large fluctuations as one approached the transition point from above; these would be driven by the nonanalyticity at the endpoint of the supercooled phase which at large $N$ would have to coincide with the first-order point.  One might also consider scenarios where something similar happens in the superheated hadronic phase.  Lattice evidence on where $T_c$ approaches $T_H$ in the large-$N$ limit is inconclusive~\cite{Bringoltz_2006}. The two temperatures are the same in $\mathcal N=4$ super Yang-Mills, as well as in some models of the large-$N$ limit of $SU(N)$ Yang-Mills~\cite{Sundborg:1999ue}.  Any scenario of this sort is worth exploring.

One very interesting possibility concerns the hadronic region with energy density beyond $\epsilon_H$.  It may turn out that in certain circumstances, this region is long-lived at large $N$, in a parametric sense, despite being locally unstable.  The basic issue is that at large $N$ the system is not unstable, it is neutral.  Thus, the instability must come about due to $1/N$ effects, which implies that the life-time of the matter before the instability substantially effects things must be long.  There is a subtlety, however, there is also a natural time scale for which such a system can thermally equilibrate.   This is also long at large $N$, since the interactions needed to equilibrate it are also $1/N$ suppressed.  Thus, there remains an interesting open question: at large $N$ does the system equilibrate more rapidly in a parametric sense than the natural time scale for the instability?  If it does, then the possibility that a homogeneous medium could form and equilibrate and, if $N$ were sufficiently large, live for a parametrically long time before the instability destroyed it --- one would have a locally unstable but nevertheless long-lived system.  At this stage it is unclear if this happens.  It is worth noting that it is possible that this may depend on whether one is studying large-$N$ QCD or pure gauge theory since the $1/N$ corrections for glueballs and mesons differ.

Lastly, it is worth noting that much of the analysis done in the paper applies to large-$N$ Yang-Mills, and thus  QCD in $2+1$ dimensions. In $2+1$ dimensions, lattice studies show that Yang-Mills with $N = 2$ and $N=3$ have second-order phase transitions, while the transition is of a weak first-order for $N = 4$~\cite{Liddle:2008kk,Holland:2007ar}. Lattice simulations suggest that the first-order transition persists at larger $N$~\cite{Liddle:2008kk,Holland:2005nd}. With Assumption~\ref{nonan}, the analysis in Section~\ref{gluon} holds for large-$N$ theories in $2+1$ dimensions. The analysis in Section~\ref{hadron} also holds as the large-$N$ QCD has the Hagedorn spectrum in $2+1$ dimensions~\cite{Cohen:2011yx}. Such rich phenomena of metastable and unstable phases of large-$N$ Yang-Mills theories in $2+1$ dimensions can be studied numerically on a lattice. 


\begin{acknowledgments}
T.C., S.L., and Y.Y. are supported by the U.S. Department of Energy under Contract No.~DE-FG02-93ER-40762. The authors thank Joseph Kapusta and Robert Pisarski for helpful suggestions.
\end{acknowledgments}

\bibliographystyle{apsrev4-1}
\bibliography{LargeN}
\end{document}